\documentclass[12pt]{article}
 \usepackage{fullpage}
\usepackage{lmodern}
\usepackage[utf8]{inputenc}
\usepackage[T1]{fontenc}

\usepackage[textsize=tiny,backgroundcolor=green!40]{todonotes}
\usepackage{booktabs}
\usepackage[colorlinks=true,allcolors=black]{hyperref}

\expandafter\let\csname equation*\endcsname\relax
\expandafter\let\csname endequation*\endcsname\relax
\usepackage{amsthm,amsmath,amssymb}

\usepackage{cleveref}
\usepackage{multicol}
\usepackage{subcaption}

\usepackage{adjustbox}
\usepackage{braket}

\usepackage{longtable}

\usepackage{tikz}
\usetikzlibrary{arrows, automata, shapes, shapes.geometric, quantikz, calc, patterns, decorations, decorations.pathmorphing, fadings}
\tikzset{elliptic state/.style={draw,ellipse}}
\usepackage{pgfplots}
\usetikzlibrary{pgfplots.groupplots}

\usepackage{listings}
\renewcommand{\emptyset}{\varnothing}

\newcommand{\QUILC}{\textsc{Quilc}}
\newcommand{\QUIL}{Quil}

\newtheorem{theorem}{Theorem}
\theoremstyle{definition}

\theoremstyle{remark}

\newtheorem{example}[theorem]{Example}

\makeatletter
\renewcommand\@date{{%
  \vspace{-\baselineskip}%
  \large\centering
Robert S.\ Smith, Eric C.\ Peterson, Mark G.\ Skilbeck, Erik J.\ Davis
  \medskip

  \textit{Rigetti Computing}\footnote{Address: 2919 Seventh St., Berkeley, CA 94710}

  \bigskip\bigskip

  \today
}}
\makeatother

\begin{document}

\title{An Open-Source, Industrial-Strength Optimizing Compiler for Quantum Programs}

\maketitle

\begin{abstract}
\QUILC{} is an open-source, optimizing compiler for gate-based quantum programs written in \QUIL{} or QASM, two popular quantum programming languages. The compiler was designed with attention toward NISQ-era quantum computers, specifically recognizing that each quantum gate has a non-negligible and often irrecoverable cost toward a program's successful execution. \QUILC{}'s primary goal is to make authoring quantum software a simpler exercise by making architectural details less burdensome to the author. Using \QUILC{} allows one to write programs faster while usually not compromising---and indeed sometimes improving---their execution fidelity on a given hardware architecture. In this paper, we describe many of the principles behind \QUILC{}'s design, and demonstrate the compiler with various examples.
\end{abstract}

\section{Introduction}

Noisy intermediate-scale quantum (NISQ) computers are an active area of research. New quantum computer architectures are sometimes the result of incremental improvements in the manufacturing process, and at other times are paradigm-shifts in the qubit technologies themselves. While each new architecture is universal in a computational sense, the impermanence of their designs challenges one's ability to write software for them. As has been the case with classical computers, the role of a compiler is to attenuate this challenge. Software for a quantum computer is ideally written in the manner that is simplest and most straightforward to the programmer, without necessarily requiring knowledge of the particulars of the target architecture. It is then the job of the compiler to produce both an efficient and an appropriate expression of this software which accounts for the details of the target architecture.

In this paper we present \QUILC{}, an open-source\footnote{The source code of \QUILC{} is licensed under the Apache~2.0 license. The source code can be found at \url{github.com/rigetti/quilc}. This document refers to \QUILC{} version~1.12.1.} software application used to compile quantum programs written in \QUIL{}~\cite{smith2016practical, QUILSPEC} into an optimized program that is expressed in the native operations of a target quantum computer architecture. \QUILC{} does not require---and indeed has no means to accept---instruction from the user on a fine-grained compilation strategy. Instead, it consumes a simple description of the architecture for which \QUILC{} must compile the user's program. The architecture description language is general enough to handle most\footnote{Currently, architectures with particularly exotic gates acting on 3+ qubits aren't well supported.} manufactured gate-based computer architectures to date, and anticipates new ones. For these reasons, we say \QUILC{} is \emph{automatic} and \emph{retargetable}.  \QUILC{} is also more than a desk calculator---a convenience to avoid doing manual, repetitive calculations---as it acts as a repository of knowledge about the compilation of programs, and it is able to synthesize this information to discover non-trivial expressions of a quantum program. We provide examples of this in \Cref{sec:examples}. It is also production-grade, and is deployed as an essential component of Rigetti Computing's software stack.

The structure of the paper is as follows.  First, in \Cref{sec:target} we provide an overview of \QUILC{}, including a mathematical formulation of quantum architectures as they pertain to compilation.  This formalism is used in \Cref{sec:structure} to describe how \QUILC{} achieves retargetablity, a high-level overview of which is presented in \Cref{QuilcStagesOverview}. In \Cref{sec:CloserLook} we consider two compilation stages (the ``addressing'' and ``compression'' stages) in more detail. We follow this in \Cref{sec:examples} with a few non-trivial examples, which make use of many of the features present in \QUILC{}. In \Cref{sec:Performance} we investigate the performance of \QUILC{} on a set of benchmarks. Finally, in the appendix we consider some implementation details which may be of interest to compiler authors or potential contributors to \QUILC{}, including examples of our ``compilation subroutine'' domain-specific language, additional features of \QUILC{}, and a short history of the project's development.

\begin{figure}
    \centering
    \begin{tabular}{ccccccccc}
    original & $\rightsquigarrow$ & layout & $\rightsquigarrow$ & nativization & $\rightsquigarrow$ & optimization \\
    \hline
    \texttt{CNOT 0 4}
    & $\rightsquigarrow$ &
    \texttt{CNOT 4 5}
    & $\rightsquigarrow$ &
    \begin{tabular}{l}
    \texttt{RX(pi/2) 5} \\
    \texttt{RZ(pi/2) 5} \\
    \texttt{RX(-pi/2) 5} \\
    \texttt{CZ 5 4} \\
    \texttt{RZ(pi) 4} \\
    \texttt{RX(pi/2) 5} \\
    \texttt{RX(0.0) 4} \\
    \texttt{RZ(-pi/2) 5} \\
    \texttt{RZ(0.0) 4} \\
    \texttt{RX(-pi/2) 5}
    \end{tabular}
    & $\rightsquigarrow$ &
    \begin{tabular}{l}
    \texttt{RZ(-pi/2) 5} \\
    \texttt{RX(pi/2) 5} \\
    \texttt{CZ 5 4} \\
    \texttt{RZ(pi) 4} \\
    \texttt{RX(-pi/2) 5} \\
    \texttt{RZ(pi/2) 5}
    \end{tabular}
    \end{tabular}
    \caption{Compilation of the program \texttt{CNOT 0 4}. The layout stage looks for an optimal initial mapping of logical qubits to physical qubits: in this example, the compiler has opted to relabel qubit 0 as qubit 4, and qubit 4 as qubit 5, so that in the (here unspecified) chip topology the two-qubit interaction can occur without introducing \texttt{SWAP} instructions. The nativization stage converts any non-native gate into a native gate; in this example \texttt{CNOT} is first compiled to \texttt{RY}, \texttt{Z}, and \texttt{CZ} gates, and then finally nativization compiles those \texttt{RY} and \texttt{Z} gates into native \texttt{RX}, \texttt{RZ} and \texttt{CZ} gates.  The final optimization stage here rewrites several \texttt{RX} or \texttt{RZ} instructions acting on the same target qubit into fewer instructions.}
    \label{QuilcStagesOverview}
\end{figure}

\section{A Quantum Compiler Target} \label{sec:target}

The structure of \QUILC{} is informed by the task at hand: it must conform to the features and constraints of gate-based quantum computational devices (i.e., the shape of its output), and it must understand the features and constraints of the quantum programming language, \QUIL{} (i.e., the shape of its input).  Gate-based quantum computational devices are made up of quantum resources (typically qubits) and support operations which affect the state of a subsystem, commingle the states of two or more subsystems, or collapse and copy the state of a subsystem for classical interpretation.  The \QUIL{} language itself provides support for all of these operations in an assembly-like format: the system's resources are addressed by ``quantum registers'', the individual assembly instructions correspond to the aforementioned operations, and there are mechanisms for further specifying other classical requirements (e.g., memory to store user-defined parameters) and their operation.

\begin{figure}
\begin{multicols}{2}
\begin{lstlisting}
DECLARE beta  REAL
DECLARE gamma REAL
DECLARE ro    BIT[3]
H 0
H 1
H 2
CPHASE(beta) 0 1
CPHASE(beta) 0 2
CPHASE(beta) 1 2



RX(gamma) 0
RX(gamma) 1
RX(gamma) 2
MEASURE 0 ro[0]
MEASURE 1 ro[1]
MEASURE 2 ro[2]
\end{lstlisting}
\end{multicols}
\caption{A typical, small instance of a ``QAOA'' quantum program.  The instructions \texttt{H}, \texttt{CPHASE}, \texttt{RX}, and \texttt{MEASURE} respectively put the qubit registers into superposition states, commingle the registers, attempt to ``unmix'' them, and measure the residual commingling.}
\label{listingqaoa}
\end{figure}

\begin{figure}
\begin{multicols}{2}
\begin{lstlisting}
DEFCIRCUIT RESET q scratch:
    MEASURE q scratch
    JUMP-UNLESS @done scratch
    X q
    LABEL @done
\end{lstlisting}
\begin{lstlisting}
DEFGATE U(%alpha, %beta):
    cis(2*pi*%alpha),  0, 0, 0
    0, cis(-2*pi*%alpha), 0, 0
    0, 0,  cis(-2*pi*%beta), 0
    0, 0, 0,   cis(2*pi*%beta)
\end{lstlisting}
\end{multicols}
\caption{On the left: a \QUIL{} snippet demonstrating the implementation of a \texttt{RESET} instruction as applied to a qubit \texttt{q}.  The instruction \texttt{X} is the quantum equivalent of a \texttt{NOT} instruction. On the right: a \QUIL{} snippet defining a custom instruction.}
\label{listingcircuitsandgates}
\end{figure}

To remain portable, the \QUIL{} language is also designed to be hardware-agnostic: it makes no particular assumptions on the availability of resources or what particular operations they support.  Instead, its execution semantics are formally specified against a mathematical backend in such a way that makes it clear how to abstractly simulate the effects of a \QUIL{} program.  At the other extreme, physical devices do labor under a host of severe constraints: there is a fixed (and typically small) number of resources, there are only a few very particular instructions which the device can enact on those resources, operations are subject to other requirements (e.g., spatial locality: distant qubits are typically unable to directly interact), and operations may be error-prone.

The role of a quantum compiler is to convert an abstract specification of such a quantum program into machine-executable bytecode, interpretable by the classical electronics which manipulate the engineered quantum system.  The compilation process cleaves roughly into two parts: the conversion of the program's quantum aspects to a form that comports with the constraints of the engineered quantum system, followed by the conversion of the classical aspects to a form that comports with the structure of the control electronics.  The quantum concerns are primarily those announced above, and while we will chiefly concern ourselves with them, we also name some classical concerns for completeness: memory management, timing and synchronization, classical communication, as well as a host of others.  \QUILC{}'s approach to the satisfaction of the quantum constraints is to order them by severity:
\begin{enumerate}
    \item Any operations of large arity must be decomposed into an arity supported by the system.\label{ArityReq}
    \item Any operations between non-interacting or indirectly interacting regions must be spatially rearranged to accommodate the system's preferred set of interactions.\label{AddressingReq}
    \item All operations must be written in terms of the set of operations (presumed universal) that the system can perform.\label{NativeGateSetReq}
    \item The program should be structured so as to avoid suffering performance penalties.\label{DepthReq}
\end{enumerate}
The first three requirements are all equally important, in the sense that the program cannot be executed on a given physical device if any are not satisfied---but on a heterogeneous device, it's not possible to discern what operations the system can perform without first resolving \eqref{AddressingReq}, which in turn requires resolving \eqref{ArityReq}, so that only then can \eqref{NativeGateSetReq} be redressed.  \QUILC{} leaves the resolution of \eqref{DepthReq} for last, since it can be satisfied ``by degrees'': for instance, it is generally better for \eqref{DepthReq} to use fewer instructions, but there are also no clear hard limits to either of the maximum count tolerable or the minimum count achievable.\footnote{The decision to resolve arity and addressing before circuit optimization does incur a loss of information. In particular, certain high-level circuit identities may not be obvious after the program has been lowered to elementary operations. However, in many instances such high-level circuit optimizations are most appropriately handled by library authors, in concert with the low-level optimizations of the compiler. In this respect, programming a quantum computer does not differ so much from programming a classical one.}

There is a natural data structure which captures these constraints of the target architecture and which forms the backbone of \QUILC{}.  To describe it, fix a set $S$ of quantum resources.  A target architecture amounts to specifying the available \textit{interacting subsets} $S' \subseteq S$, each indicating a collection of resources that are permitted to interact---for instance, a pair of qubits on which one can perform a two-qubit gate.  Let us make the additional assumption that if $S'$ is an interacting subset, then every further nonempty subset $S'' \subseteq S'$ is also an interacting subset.  Under this assumption, such a collection of interacting subsets forms a simplicial complex $\Sigma$~\cite[Section 3.1]{Spanier} with $S$ as its set of vertices (or $0$--simplices), interacting pairs as its edges (or $1$--simplices), interacting triplets as its $2$--simplices, and so on.  Additionally, we tag each simplex with a set of \textit{instructions}, each corresponding to a specific means by which the interacting set can evolve as an ensemble.  Finally, each such instruction is tagged with metadata, e.g., a matrix defining the associated unitary transformation, the average fidelity of the device's execution of the instruction compared to the ideal, or the temporal duration of the instruction execution on the device.

We have arranged the description of the target architecture into these tiers because this reflects the kind of information needed as input to the four compilation constraints:
\begin{enumerate}
    \item The limit on the arity of an instruction corresponds to the dimension of the largest simplex.
    \item The simplicial complex structure of interactions describes both the spatial constraints to which a quantum program is subject, as well as the pathways by which information can be rerouted or permuted in order to satisfy these constraints.
    \item The instruction tags indicate into which gateset the program's components must be compiled.
    \item Metadata can be used to make decisions that boost overall program performance.  For instance, fidelity information might prefer one region of the device over another, or a particular instruction decomposition over another.
\end{enumerate}

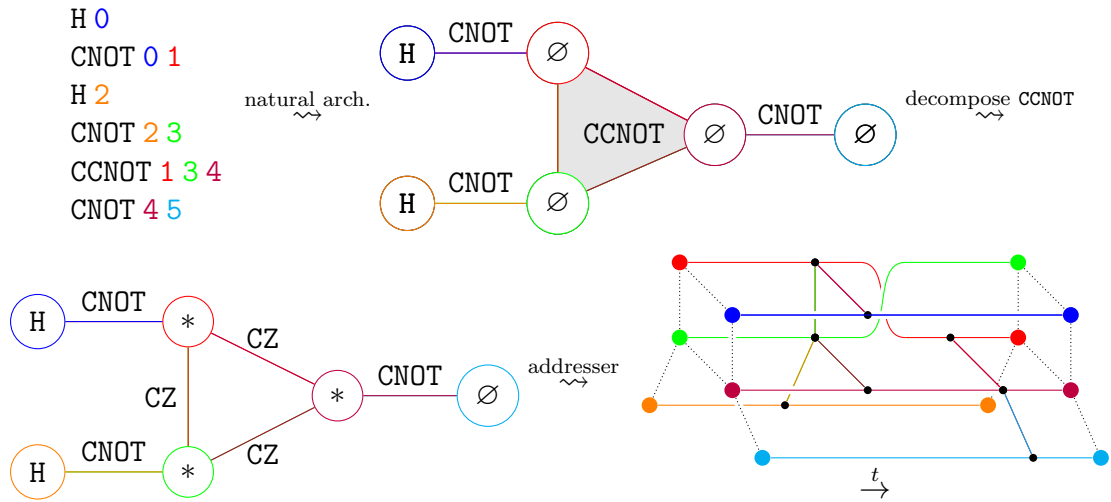
\begin{figure}
    \[
    \begin{array}{l}
        \mathtt{H\:\textcolor{blue}{0}} \\
        \mathtt{CNOT\:\textcolor{blue}{0}\:\textcolor{red}{1}} \\
        \mathtt{H\:\textcolor{orange}{2}} \\
        \mathtt{CNOT\:\textcolor{orange}{2}\:\textcolor{green}{3}} \\
        \mathtt{CCNOT\:\textcolor{red}{1}\:\textcolor{green}{3}\:\textcolor{purple}{4}} \\
        \mathtt{CNOT\:\textcolor{purple}{4}\:\textcolor{cyan}{5}} \\
    \end{array}
    \overset{\text{natural arch.}}\rightsquigarrow
    \begin{tikzpicture}[node distance=2cm,baseline=(current bounding box.center)]
        \node[shape=circle,draw=black] (A) {\texttt{H}};
        \node[shape=circle,draw=black,right of=A] (B) {$\emptyset$};
        \node[shape=circle,draw=black,below of=A] (C) {\texttt{H}};
        \node[shape=circle,draw=black,right of=C] (D) {$\emptyset$};
        \node[shape=circle,draw=black,below right=0.5cm and 1.5cm of B] (E) {$\emptyset$};
        \node[shape=circle,draw=black,right of=E] (F) {$\emptyset$};
        
        \path [-](A) edge[path fading=west, draw=red] (B);
        \path [-](A) edge[path fading=east, draw=blue] (B);
        \path(A) -- node[above]{$\mathtt{CNOT}$} (B);
        
        \path [-](C) edge[path fading=west, draw=green] (D);
        \path [-](C) edge[path fading=east, draw=orange] (D);
        \path (C) -- node[above]{$\mathtt{CNOT}$} (D);
        
        \path [-](B) edge[path fading=east, draw=red] (E);
        \path [-](B) edge[path fading=west, draw=purple] (E);
        
        \path [-](D) edge[path fading=north, draw=green] (E);
        \path [-](D) edge[path fading=south, draw=purple] (E);
        
        \path [-](B) edge[path fading=north, draw=green] (D);
        \path [-](B) edge[path fading=south, draw=red] (D);
        
        \path [-](E) edge[path fading=west, draw=cyan] (F);
        \path [-](E) edge[path fading=east, draw=purple] (F);
        \path (E) -- node[above]{$\mathtt{CNOT}$} (F);

        \fill[fill=gray!20] (B.center)--(D.center)--(E.center)--cycle;
        
        \path [-](B) edge[path fading=east, draw=red] (E);
        \path [-](B) edge[path fading=west, draw=purple] (E);
        
        \path [-](D) edge[path fading=north, draw=green] (E);
        \path [-](D) edge[path fading=south, draw=purple] (E);
        
        \path [-](B) edge[path fading=north, draw=green] (D);
        \path [-](B) edge[path fading=south, draw=red] (D);
        
        \node[shape=circle,draw=blue] (A) {\texttt{H}};
        \node[shape=circle,draw=red,right of=A,fill=white] (B) {$\emptyset$};
        \node[shape=circle,draw=orange,below of=A] (C) {\texttt{H}};
        \node[shape=circle,draw=green,right of=C,fill=white] (D) {$\emptyset$};
        \node[shape=circle,draw=purple,below right=0.5cm and 1.5cm of B,fill=white] (E) {$\emptyset$};
        \node[shape=circle,draw=cyan,right of=E] (F) {$\emptyset$};
        
        \node[below right=0.5cm and -0.1cm of B] (CCNOT) {$\mathtt{CCNOT}$};
    \end{tikzpicture}
    \overset{\text{decompose \texttt{CCNOT}}}\rightsquigarrow
    \]
    \[
    \begin{tikzpicture}[node distance=2cm,baseline=(current bounding box.center)]
        \node[shape=circle,draw=blue] (A) {\texttt{H}};
        \node[shape=circle,draw=red,right of=A] (B) {$*$};
        \node[shape=circle,draw=orange,below of=A] (C) {\texttt{H}};
        \node[shape=circle,draw=green,right of=C] (D) {$*$};
        \node[shape=circle,draw=purple,below right=0.5cm and 1.5cm of B] (E) {$*$};
        \node[shape=circle,draw=cyan,right of=E] (F) {$\emptyset$};
    
        \path [-](A) edge[path fading=west, draw=red] (B);
        \path [-](A) edge[path fading=east, draw=blue] (B);
        \path(A) -- node[above]{$\mathtt{CNOT}$} (B);
        
        \path [-](C) edge[path fading=west, draw=green] (D);
        \path [-](C) edge[path fading=east, draw=orange] (D);
        \path (C) -- node[above]{$\mathtt{CNOT}$} (D);
        
        \path [-](B) edge[path fading=east, draw=red] (E);
        \path [-](B) edge[path fading=west, draw=purple] (E);
        \path (B) -- node[above]{$\mathtt{CZ}$} (E);
        
        \path [-](D) edge[path fading=north, draw=green] (E);
        \path [-](D) edge[path fading=south, draw=purple] (E);
        \path (D) -- node[below]{$\mathtt{CZ}$} (E);
        
        \path [-](B) edge[path fading=north, draw=green] (D);
        \path [-](B) edge[path fading=south, draw=red] (D);
        \path (B) -- node[left]{$\mathtt{CZ}$} (D);
        
        \path [-](E) edge[path fading=west, draw=cyan] (F);
        \path [-](E) edge[path fading=east, draw=purple] (F);
        \path (E) -- node[above]{$\mathtt{CNOT}$} (F);
    \end{tikzpicture}
    \overset{\text{addresser}}\rightsquigarrow
\begin{array}{c}
\begin{tikzpicture}[baseline=(current bounding box.center)]
\newlength{\step}
\setlength{\step}{0.6cm}
\newlength{\rad}
\setlength{\rad}{3pt}
\tikzset{every node/.style={circle,fill=black,inner sep=0pt,minimum size=\rad}}
\tikzset{
  crossing over/.style={
    /tikz/preaction={
      /tikz/draw=white,
      /tikz/arrows=-,
      /tikz/line width=3pt}}}

\node[red, minimum size=2\rad] (A) {};
\node[green, minimum size=2\rad] (B) at ([yshift=-1cm]A) {};
\node[blue, minimum size=2\rad] (C) at ([shift=({0.7cm,-0.7cm})]A) {};
\node[orange, minimum size=2\rad] (D) at ([shift=({-0.4cm,-0.9cm})]B) {};
\node[purple, minimum size=2\rad] (E) at ([yshift=-1cm]C) {};
\node[cyan, minimum size=2\rad] (F) at ([shift=({0.4cm,-0.9cm})]E) {};

\node[red,    minimum size=2\rad] (AR) at ([xshift=7.5\step]B) {};
\node[green,  minimum size=2\rad] (BR) at ([xshift=7.5\step]A) {};
\node[blue,   minimum size=2\rad] (CR) at ([xshift=7.5\step]C) {};
\node[orange, minimum size=2\rad] (DR) at ([xshift=7.5\step]D) {};
\node[purple, minimum size=2\rad] (ER) at ([xshift=7.5\step]E) {};
\node[cyan,   minimum size=2\rad] (FR) at ([xshift=7.5\step]F) {};

\draw[densely dotted] (DR) -- (AR) -- (ER) -- (FR);
\draw[densely dotted] (AR) -- (BR) -- (CR) -- (ER);

\coordinate (SWAP3Source) at ([xshift=4\step]A);
\coordinate (SWAP3Target) at ([xshift=5\step]B);
\coordinate (SWAP4Source) at ([xshift=4\step]B);
\coordinate (SWAP4Target) at ([xshift=5\step]A);
\draw[red] (A) -- (SWAP3Source);
\draw[red] (SWAP3Source) to[out=0,in=180] (SWAP3Target);
\draw[green] (B) -- (SWAP4Source);
\draw[green, crossing over] (SWAP4Source) to[out=0,in=180] (SWAP4Target);
\draw[red] (SWAP3Target) -- (AR);
\draw[green] (SWAP4Target) -- (BR);
\draw[orange] (D) -- (DR);
\draw[cyan] (F) -- (FR);

\draw[densely dotted, crossing over] (D) -- (B) -- (E) -- (F);
\draw[densely dotted, crossing over] (B) -- (A) -- (C) -- (E);

\node[circle,fill=black,minimum size=\rad] (CNOT23L) at ([xshift=3\step]D) {};
\node[circle,fill=black,minimum size=\rad] (CNOT23R) at ([xshift=3\step]B) {};
\draw[path fading=south, green] (CNOT23L) -- (CNOT23R);
\draw[path fading=north, orange] (CNOT23L) -- (CNOT23R);

\draw[purple, crossing over] (E) -- (ER);

\node[circle,fill=black,minimum size=\rad] (CNOT13L) at ([xshift=3\step]A) {};
\node[circle,fill=black,minimum size=\rad] (CNOT13R) at ([xshift=3\step]B) {};
\draw[path fading=south, red] (CNOT13L) -- (CNOT13R);
\draw[path fading=north, green] (CNOT13L) -- (CNOT13R);

\node[circle,fill=black,minimum size=\rad] (CNOT34L) at ([xshift=3\step]B) {};
\node[circle,fill=black,minimum size=\rad] (CNOT34R) at ([xshift=3\step]E) {};
\draw[path fading=south, green] (CNOT34L) -- (CNOT34R);
\draw[path fading=north, purple] (CNOT34L) -- (CNOT34R);

\node[circle,fill=black,minimum size=\rad] (CNOT14L) at ([xshift=6\step]B) {};
\node[circle,fill=black,minimum size=\rad] (CNOT14R) at ([xshift=6\step]E) {};
\draw[path fading=south, red] (CNOT14L) -- (CNOT14R);
\draw[path fading=north, purple] (CNOT14L) -- (CNOT14R);

\node[circle,fill=black,minimum size=\rad] (CNOT45L) at ([xshift=6\step]E) {};
\node[circle,fill=black,minimum size=\rad] (CNOT45R) at ([xshift=6\step]F) {};
\draw[crossing over, path fading=south, purple] (CNOT45L) -- (CNOT45R);
\draw[path fading=north, cyan] (CNOT45L) -- (CNOT45R);

\draw[blue, crossing over] (C) -- (CR);

\node[circle,fill=black,minimum size=\rad] (CNOT01L) at ([xshift=3\step]A) {};
\node[circle,fill=black,minimum size=\rad] (CNOT01R) at ([xshift=3\step]C) {};
\draw[crossing over, path fading=north, blue] (CNOT01L) -- (CNOT01R);
\draw[path fading=south, red] (CNOT01L) -- (CNOT01R);

\draw[blue] (C) -- (CR);

\end{tikzpicture} \\ \stackrel{t}{\to} \end{array}
    \]
    \caption{An example input program, its associated natural architecture, its native architecture after decomposing the $\mathtt{CCNOT}$ into two-qubit interactions, and the effect of the addresser (cf.\ \Cref{sec:structure}, only $\mathtt{CZ}$s displayed) when targeting a particular device shaped like a square with two arms.}
    \label{SigmasFigure}
\end{figure}

\begin{figure}
\begin{lstlisting}[language=c]
{"1Q": {
  "0": {
   "gates": [
    {"operator": "RX", "parameters": [1.5707963267948966], "arguments": [0]},
    {"operator": "RZ", "parameters": ["_"],                "arguments": [0]}]},
  "1": {
   "gates": [
    {"operator": "RX", "parameters": [1.5707963267948966], "arguments": [0]},
    {"operator": "RZ", "parameters": ["_"],                "arguments": [0]}]}},
 "2Q": {
  "0-1": {
   "gates": [
    {"operator": "CZ", "parameters": [],                   "arguments": [0, 1]}]}}}
\end{lstlisting}
\caption{A simple target architecture in serialized form. Note that $\pi/2 = 1.57\ldots$.}
\label{JSONArchitecture}
\end{figure}

As an example, we've included in \Cref{SigmasFigure} an example input program, an architecture $\Sigma$ for which it is fully native, and an example target architecture. As a further simple example, we include in \Cref{JSONArchitecture} a serialized such target architecture (without any extraneous metadata, like instruction fidelities).

\section{The Large-Scale Structure of \QUILC{}} \label{sec:structure}

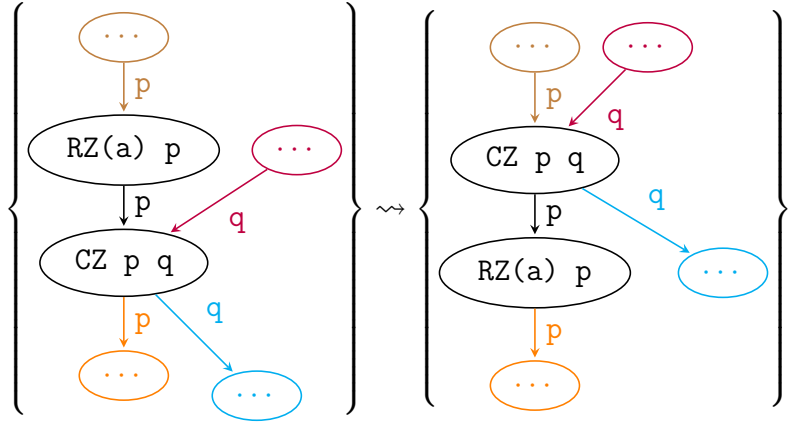
\begin{figure}
\[\left\{
\begin{tikzpicture}[baseline=(current bounding box.center),
            > = stealth, 
            shorten > = 1pt, 
            auto,
            node distance = 1.5cm, 
            semithick 
        ]
        
        \node[elliptic state] (a) {\texttt{RZ(a) p}};
        \node[elliptic state, purple] (b) [right of=a, node distance=2.3cm] {$\cdots$};
        
        \node[elliptic state, brown] (adots) [above of=a] {$\cdots$};
        
        \node[elliptic state] (CZ) [below of=a] {\texttt{CZ p q}};
        
        \node[elliptic state, orange] (CZdots) [below of=CZ] {$\cdots$};
        \node[elliptic state, cyan] (c) [below right of=CZ, node distance=2.5cm] {$\cdots$};
        
        \path[->, brown] (adots) edge node {\texttt{p}} (a);
        \path[->] (a) edge node {\texttt{p}} (CZ);
        \path[->, purple] (b) edge node {\texttt{q}} (CZ);
        \path[->, orange] (CZ) edge node {\texttt{p}} (CZdots);
        \path[->, cyan] (CZ) edge node {\texttt{q}} (c);
\end{tikzpicture}
\right\} \rightsquigarrow \left\{
\begin{tikzpicture}[baseline=(current bounding box.center),
            > = stealth, 
            shorten > = 1pt, 
            auto,
            node distance = 1.5cm, 
            semithick 
        ]
        
        \node[elliptic state, brown] (CZdots) {$\cdots$};
        \node[elliptic state, purple] (b) [right of=CZdots] {$\cdots$};
        
        \node[elliptic state] (CZ) [below of=CZdots] {\texttt{CZ p q}};
        
        \node[elliptic state] (a) [below of=CZ] {\texttt{RZ(a) p}};
        \node[elliptic state, cyan] (c) [right of=a, node distance=2.5cm] {$\cdots$};
        
        \node[elliptic state, orange] (adots) [below of=a] {$\cdots$};
        
        \path[->, brown] (CZdots) edge node{\texttt{p}} (CZ);
        \path[->, purple] (b) edge node{\texttt{q}} (CZ);
        \path[->] (CZ) edge node{\texttt{p}} (a);
        \path[->, cyan] (CZ) edge node{\texttt{q}} (c);
        \path[->, orange] (a) edge node{\texttt{p}} (adots);
\end{tikzpicture}\right\}\]
    \caption{An example graph-based rewriting rule, commuting an \texttt{RZ} past a \texttt{CZ}.  The vertices are labeled by instructions, the edges are labeled by the resource shared by the source and target, and the direction indicates which instruction precedes the other in the program.  The black region consists of vertices which are matched, destroyed, and recreated; the various colored regions are detached from the old black vertices and reattached to the new black vertices as indicated.}
    \label{RewritingRuleExample}
\end{figure}

Let us now turn to our high-level overview of \QUILC{}.  \QUILC{} segments the satisfaction of the above constraints into the following stages:
\begin{description}
    \item[Lexing/parsing] First, \QUILC{} parses plain-text input into a syntax tree.
    \item[Control-flow graph construction] It then segments the input program into classical and non-classical components, so that consideration of the effects of classical instructions can be deferred to the compiler backend.  The primary mechanism for ``hiding'' these instructions (e.g., jumps) is to store them as labels in a control flow graph.
    \item[Addressing] \QUILC{} resolves constraints \eqref{ArityReq}, \eqref{AddressingReq}, and \eqref{NativeGateSetReq} by walking the graph of quantum instructions, ordered by resource dependence, in a breadth-first manner and tracking a mapping from logical quantum resources (i.e., those specified in the user's program) to physical quantum resources (i.e., those actually available on the device).
    \item[Compression] Lastly, \QUILC{} gives attention to \eqref{DepthReq} by finding sequences of instructions that act on overlapping resources through a kind of depth-first walk and applying reduction techniques (e.g., a peephole rewriter) to the paths appearing during the walk.  See \Cref{RewritingRuleExample} for an example.
\end{description}

There are two important points to note.  First, the addressing and compression stages require intensive computation to fully explore the graph: for example, deciding whether the addressing problem admits a solution without the insertion of \texttt{SWAP} instructions is an instance of the (NP-complete) subgraph isomorphism problem~\cite{siraichi2018qubit}.  Instead, we employ heuristics, randomization, and approximate algorithms to tame their computational complexity; see \Cref{sec:CloserLook} for more details, as well as  \Cref{BenchmarkFigure} and \Cref{QFTBenchmarkFigure} in \Cref{ImpDetailsSection} for some empirical analysis of these approximations.  The second important point is that although the addressing and compressing stages crawl the program in a highly different way, the actual manipulation of quantum instructions is similar in each each. This promotes the segmentation of these stages into small, interruptible \textit{compilation subroutines} which adopt a uniform interface.  This API which we expect compilation subroutines to provide consists of
\begin{itemize}
    \item A literal subroutine, which consumes some fixed number of instructions and some data about the state of the compiler, and which emits a sequence of instructions which may replace its input.  The subroutine is allowed to be partially defined: compilation subroutines can signal that they are not applicable in a given situation by employing an interrupt which is handled appropriately by the caller.
    \item A description of the kinds of instructions that the subroutine can consume.
    \item A description of the kinds of instructions (and, ideally, their counts) that the routine can emit.
\end{itemize}
Equipped with this extra data, \QUILC{} can decide whether a particular subroutine falls into one of two privileged classes (or neither):
\begin{description}
    \item[Nativizers] A compilation routine is relevant for nativization when it consumes a single instruction and when its output belongs to the native interactions of $\Sigma$ (perhaps after further applications of other nativizers).
    \item[Optimizers] A compilation routine is relevant for optimization when it both consumes and emits instruction sequences which belong to the native interactions of $\Sigma$ and its emitted sequences have better execution properties than its inputs.
\end{description}
The addressing step employs the first class of compilation subroutines in order to convert the input program's non-native instructions to instructions that are native for $\Sigma$'s interactions.  The compression step makes use of both of these special classes: the optimizers are directly relevant to the compression of instruction sequences, but it is also fruitful to destroy some of the structure of a sequence of instructions by considering their holistic effect and to re-nativize it.

This kind of design lends itself to the support of a few features:
\begin{description}
    \item[Planning] Verifying that a particular long sequence of reductions gives an optimal strategy (from the perspective of constraint \eqref{DepthReq}) is computationally expensive, frequently remitted to heuristic, and costly to guess incorrectly.  It is less expensive to make analogous decisions about individual, simple reductions, and so interruptibility permits the compiler a greater degree of flexibility in quickly planning its next best move.
    \item[Internal reusability] Requiring that subroutines leave the overall compiler in a good intermediate state puts dramatic limitations on the API to which they conform.  From the perspective of \QUILC{}, this tends to make such subroutines suitable for use at various stages of compilation.  It also promotes a separation of concerns between the code responsible for crawling the input program (as described above in, e.g., ``addressing'' and ``compression'') and these subroutines, so that the crawlers are written in such a generic way that their specification does not directly depend on knowing the set of subroutines to be applied to the input program.
    \item[Ease of authorship] The same limitations on the API means that the author of quantum-specific subroutines need not concern themselves with the precise implementation of the crawlers---or, in \QUILC{}'s case, even how to instruct the crawlers that they should make use of a new subroutine.
    \item[External reusability] It is possible to wrap routines provided by external compilation libraries via this API, so that \QUILC{} can make seamless use of their specialized routines without reimplementation or serious diversion. One such example is \texttt{tweedledum}, a quantum circuit optimizing library which provides an efficient routine for compilation of gates that can be represented as permutation matrices~\cite{tweedledum}.
\end{description}


\section{A Closer Look at Addressing and Compression}\label{sec:CloserLook}

Much of the ``action'' of the compiler occurs in the addressing and compression stages. Indeed, these are the most computationally intensive stages of compilation, and the techniques employed are critically responsible for the quality of the output of the compiler (e.g., with respect to gate depth). In this section we give brief summaries of the underlying techniques of these two compilation stages, along with points of contact in the literature.

\subsection*{Addressing}

In principle, the addressing stage may be resolved by any sequence of circuit transformations which preserve the semantics of the source program and which result in an output program conforming to the constraints of the hardware. However, the set of such possible transformations is immense. Here we constrain ourselves to two classes of transformations: those which serve to translate gates into native constituents (e.g., the aforementioned ``nativization'' routines) and those which manage the assignment of logical to physical qubits (known as ``qubit allocation'' in the literature).

A number of general-purpose techniques for qubit allocation have been proposed; we briefly mention some here. In \cite{siraichi2018qubit}, the authors consider four sorts of program transformations (``virtual \texttt{CNOT}s'', reversals, bridges, and swaps), formulating the problem precisely in these terms and presenting heuristics to approximate the optimal sequence of transformations to solve the qubit allocation problem.

In \cite{zulehner2018efficient}, the authors first produce an initial decomposition of the source circuit (consisting of 1Q and 2Q gates) into layers, then construct an initial qubit mapping, and then finally identify swap operations between layers via A* search using a cost function which may incorporate look-ahead to successive layers. The method of \cite{cowtan2019qubit} uses a similar layered decomposition, but with different heuristics for constructing the initial mapping and for selecting swaps between layers.

Alternative approaches have formulated the addressing problem in a form amenable to solution by off-the-shelf solvers. For example, in \cite{murali2019noise}, the authors formulate qubit allocation as a satisfiability modulo theories (SMT) problem which may be solved by an SMT solver. Along similar lines, \cite{venturelli2018compiling} proposes and evaluates a formulation which is solvable by temporal planners.

Core to the approach taken by \QUILC{} is a representation of the source program which expresses the constraints, with respect to hardware usage, implicit in the linear source program. Here, the source instructions are taken as the vertices of a directed, acyclic graph, with edges expressing resource conflicts (whether classical or quantum) between logically successive operations. The addressing pass proceeds in a greedy fashion, consuming the source program in topological order while maintaining a certain amount of state, including a partial logical-to-physical qubit mapping, estimated swap costs between pairs of qubits, and a buffer of emitted instructions operating on physical qubits. 

In this scheme, gate applications involving a number of qubits exceeding the underlying arity of the device (e.g., 3Q gates for Rigetti’s hardware) are first translated to an equivalent series of smaller-arity operations by means of any number of nativization routines, as described in \Cref{sec:structure}. We note that the goal here is not to find an ``optimal'' sequence of native operations, but rather to quickly find a viable realization of the gate in native terms. However, \QUILC{} does attempt to select the translation so as to be in harmony with the particular qubits' native gate sets: for instance, \texttt{ISWAP}--based decompositions are preferred to \texttt{CNOT}--based decompositions when the native gate sets includes the \texttt{ISWAP} gate and not \texttt{CNOT}.

As each low-arity operation (e.g., 1Q and 2Q gates on Rigetti’s hardware) is processed, the logical-to-physical qubit mapping may be updated, either by assigning a logical qubit to a currently unassigned physical qubit, or via the introduction of SWAP operations in order to satisfy addressing constraints. At each such decision point, the ambition of \QUILC{} is to select the action which minimizes the total cost of the final scheduled program. 

To this end, \QUILC{} employs heuristics along two axes. The first are \emph{cost heuristics}, which, given a logical \QUIL{} program and a compilation target, determine a cost indicating the ``badness'' of the program on the underlying hardware. At present, there are two of these available: a duration-based heuristic, informed by the underlying gate times of the target architecture, and a fidelity-based heuristic, informed by the reported gate fidelities of the target architecture. The second axis consists of \emph{search heuristics}, which are used to select from available swap operations in order to assign a logical gate to physical qubits in a cost-minimizing fashion. These include both A*~search as well as greedy search heuristics.

A challenge with such an approach is that \texttt{SWAP}s inserted cheaply at one point in a program may end up being costly with respect to later operations. To account for this, both the duration-based and the fidelity-based cost heuristics incorporate a look-ahead: here the cost associated with a \QUIL{} program depends not just on the next instructions, but also more weakly on those following them (via an exponential discounting factor). Here, the motivation is to dampen the miserliness of the (otherwise greedy) addressing strategy.

\subsection*{Compression}

During the compression stage, \QUILC{} employs two kinds of rewriting strategies:
\begin{enumerate}
    \item Directly apply a peephole rewriter to an instruction sequence.
    \item Convert a (pure quantum) instruction sequence to a composite matrix, rewrite the matrix as native instructions, and apply a peephole rewriter to the resulting sequence.
\end{enumerate}
Both of these are best served by sequences with two properties:
\begin{description}
    \item[Lengthiness] Long sequences provide more opportunities for the rewriter to act and give the bounded nativization routines better odds of producing a shorter sequence.
    \item[Resource-sparsity] Sequences which act only on a few resources have correspondingly fewer false negatives in the form of instructions which are nonadjacent in the sequence but which could be commuted next to one another.
\end{description}
With this in mind, the compressor has been designed to produce contiguous sequences of instructions with these properties.

The compressor first arranges the instructions in a program into a dependency graph by resource usage. It then begins forming subgraphs, tagged by resource utilization, and peforms a topological walk of the instruction graph while adhering to the following rules:
\begin{itemize}
    \item If this instruction's resources do not meet those of any subgraph and it does not contain a forbidden resource, start a new subgraph containing this instruction, and restart consideration of the next instruction in the walk.
    \item Otherwise, this instruction's resources meet one or more existing subgraphs or are forbidden. Compute the sum of their resource tags with this instruction's resources.
    \item If the resource sum contains a forbidden resource collection \emph{or} if the resource sum is larger than the compressor's limit, then do nothing with this instruction for now. Remove each met subgraph from the overall graph. For each met subgraph:
        \begin{itemize}
            \item If the resource sum contains a prohibited resource collection, then mark this subgraph's resource tag as forbidden. If the resource sum is larger than the compressor's limit, then mark the sum as forbidden.
            \item Linearize the contents of the subgraph into a sequence, and pass that sequence to the peephole rewriter.
            \item Re-walk the instructions emitted by the peephole rewriter (i.e., try to form subgraphs out of them).
            \item Unmark any forbidden resources added in this step.
        \end{itemize}
        Write this instruction, as well as any instructions in any met subgraphs (which may be newly formed, as in the above loop), out to the end results.
    \item Otherwise, merge the subgraphs which meet this instruction, add it to the newly formed subgraph, tag the subgraph with the resource sum, and proceed to the next instruction in the walk.
\end{itemize}

This walk is similar in effect to the walk considered in Iten, Soetter, and Werner~\cite{iten2019efficient}.  However, ours is somewhat less thorough, since it separately walks the graph and applies template rewriting, and since it does not perform backward matching.  It partially makes up for these deficiencies in its output by its simplicity of implementation.

In the course of the compressor's graph-walking, a given instruction may be considered by the peephole rewriter as many times as there are subresources which contain the instruction and which are contained by the subgraph's tag.  By installing a limit to the size of subgraph tags which the compressor will consider, this value becomes bounded.  In practice, even a small such limit (e.g., less than four qubits) has good run time properties without appreciable decline in output quality.

\section{Long-Form Examples}\label{sec:examples}

In this section, we demonstrate the above considerations via a few practical examples which highlight the influence of target architecture and the ubiquitous role of compilation subroutines in the compilation process. The first two subsections consider compilation of the Toffoli gate, a well studied example which has been implemented on a variety of architectures and for which optimal decompositions are known~\cite{shende2008cnot}. Our aim here is to demonstrate how various compilation routines may be combined to realize \texttt{CCNOT} across several device architectures. In the third subsection, we consider state-aware compilation applied to an example from computational chemistry.

\subsection*{SWAP recombination with different targets}

One of the basic tasks of the ``addressing'' stage of the compiler is to construct a mapping from logical qubits, as expressed in the source program, to physical qubits, as realized in a specific architecture. A guiding principle here is that constraints in physical qubit connectivity may be satisfied through the addition of appropriate \texttt{SWAP} instructions.  This introduction of \texttt{SWAP} gates comes at a price: namely, an increase in the total number of logical operations performed. This is further complicated by the demands of nativization, since for many architectures of interest \texttt{SWAP}s must be translated to native operations. Bounds for the complexity of the resulting native instruction sequence have been considered in the literature. For example, it is known that \texttt{SWAP} requires 3~\texttt{CNOT} gates~\cite{vatan2004optimal}. On the other hand, an arbitrary two-qubit unitary operator is equivalent, up to a global phase factor, to one expressed as a circuit with at most 3~\texttt{CNOT} operations~\cite{shende2004minimal}. 

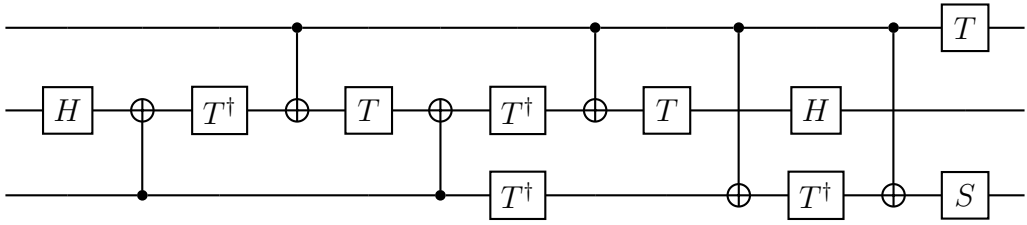
\begin{figure}
    \centering
    \begin{tikzcd}
        \qw & \qw & \qw & \qw & \ctrl{1} & \qw & \qw & \qw & \ctrl{1} & \qw & \ctrl{2} & \qw & \ctrl{2} & \gate{T} & \qw \\
        \qw & \gate{H} & \targ{} & \gate{T^{\dagger}} & \targ{} & \gate{T} & \targ{} & \gate{T^{\dagger}} & \targ{} & \gate{T} & \qw & \gate{H} & \qw & \qw & \qw \\
        \qw & \qw & \ctrl{-1} & \qw & \qw & \qw & \ctrl{-1} & \gate{T^{\dagger}} & \qw & \qw & \targ{} & \gate{T^{\dagger}} & \targ{} & \gate{S} & \qw
\end{tikzcd}
    \caption{\texttt{CCNOT} on a fully-connected chip.}
    \label{fig:connected_ccnot}
\end{figure}

Thus, for many architectures of interest, the demand of nativization presents itself as an opportunity when selecting \texttt{SWAP} targets, since the native gate cost of a single \texttt{SWAP} gate is the same as that of any subcircuit consisting of the \texttt{SWAP} gate and adjacent gates, if these are on the swapped qubits. For example, \Cref{fig:connected_ccnot} shows one realization of a \texttt{CCNOT} gate on a fully-connected chip, which is optimal in the sense that involves a total of 6 \texttt{CNOT} operations~\cite{shende2008cnot}. If nearest-neighbor connectivity is imposed, then there is a natural decision of where a \texttt{SWAP} should be inserted. Considering that nativization can convert any two-qubit unitary subcircuit into an equivalent one involving at most 3 \texttt{CNOT} gates, placing a \texttt{SWAP} on the first two qubit lines (cf.\ \Cref{fig:swap01_ccnot}) is preferred to placing on the second two lines (cf.\ \Cref{fig:swap12_ccnot}), as this is cheaper by a \texttt{CNOT} gate.

The \QUILC{} addresser makes use of this information, and more, when selecting \texttt{SWAP} placements. In practice, we observe that this ``\texttt{SWAP} recombination'' trick is compatible with additional optimizations. For example, when compiling \texttt{CCNOT 0 1 2} to a chip supporting only nearest-neighbor connectivity, \QUILC{} is able to produce native circuits containing only 7~\texttt{CNOT} gates, which is less than the 8 suggested by \Cref{fig:swap01_ccnot}.

\begin{figure}
    \centering
    \begin{tikzcd}
    \qw & \qw & \qw & \qw & \ctrl{1} & \qw & \qw & \qw \gategroup[2,steps=4,style={dashed, rounded corners,fill=blue!20, inner xsep=2pt}, background]{} & \ctrl{1} & \swap{1} & \gate{T} & \gate{H} & \gate{T} & \qw & \qw & \qw \\
    \qw & \gate{H} & \targ{} & \gate{T^{\dagger}} & \targ{} & \gate{T} & \targ{} & \gate{T^{\dagger}} & \targ{} & \targX{} & \qw & \ctrl{1} & \qw & \ctrl{1} & \qw & \qw \\
    \qw & \qw & \ctrl{-1} & \qw & \qw & \qw & \ctrl{-1} & \gate{T^{\dagger}} & \qw & \qw & \qw & \targ{} & \gate{T^{\dagger}} & \targ{} & \gate{S} & \qw
    \end{tikzcd}
    \caption{\texttt{CCNOT} with nearest-neighbor connectivity, placing a \texttt{SWAP} on the top two qubit lines. The entire highlighted subcircuit may be translated to at most 3 \texttt{CNOT} gates.}
    \label{fig:swap01_ccnot}
\end{figure}

\begin{figure}
    \centering
    \begin{tikzcd}
        \qw & \qw & \qw & \qw & \ctrl{1} & \qw & \qw & \qw & \ctrl{1} & \qw & \qw & \qw & \ctrl{1} & \qw & \ctrl{1} & \gate{T} & \qw \\
        \qw & \gate{H} & \targ{} & \gate{T^{\dagger}} & \targ{} & \gate{T} & \targ{} & \gate{T^{\dagger}} & \targ{} & \swap{1} \gategroup[2,steps=3,style={dashed, rounded corners,fill=blue!20, inner xsep=2pt}, background]{} & \gate{T^{\dagger}} & \qw & \targ{} & \gate{T^{\dagger}} & \targ{} & \gate{S} & \qw \\
        \qw & \qw & \ctrl{-1} & \qw & \qw & \qw & \ctrl{-1} & \qw & \qw & \targX{} & \gate{T} & \gate{H} & \qw & \qw & \qw & \qw & \qw
    \end{tikzcd}
    \caption{\texttt{CCNOT} with nearest-neighbor connectivity, placing a \texttt{SWAP} on the bottom two qubit lines. The entire highlighted subcircuit may be translated to at most 3 \texttt{CNOT} gates.}
    \label{fig:swap12_ccnot}
\end{figure}

\subsection*{Native targets for \texttt{CCNOT}}

\begin{figure}
\begin{multicols}{2}
\begin{lstlisting}[language=lisp]
(define-compiler CCNOT-to-CNOT
    ((input ("CCNOT" () q0 q1 q2)))
  (inst "H"    ()       q2)
  (inst "CNOT" ()       q1 q2)
  (inst "RZ"   '(-pi/4) q2)
  (inst "CNOT" ()       q0 q2)
  (inst "RZ"   '( pi/4) q2)
  (inst "CNOT" ()       q1 q2)
  (inst "RZ"   '(-pi/4) q2)
  (inst "CNOT" ()       q0 q2)
  (inst "RZ"   '( pi/4) q1)
  (inst "RZ"   '( pi/4) q2)
  (inst "CNOT" ()       q0 q1)
  (inst "H"    ()       q2)
  (inst "RZ"   '( pi/4) q0)
  (inst "RZ"   '(-pi/4) q1)
  (inst "CNOT" ()       q0 q1))
\end{lstlisting}
\end{multicols}
\caption{Compilation subroutine for implementing \texttt{CCNOT} via \texttt{CNOT}. See \Cref{sec:lisp} and \ref{sec:defcomp} for a description of the syntax.}\label{fig:ccnot_to_cnot}
\end{figure}

One of the guiding philosophies of \QUILC{} is that the burden of deciding whether a given compilation subroutine should be preferred to another in some specific context need not be borne by the quantum programmer. Instead, what is specified by the user is a set of hardware constraints.\footnote{Users do have the option of providing hints to the compiler about properties it might exploit, e.g., via command line arguments to \texttt{quilc} executable, or in \QUIL{} programs via \texttt{PRAGMA} directives.} Presented with this information, \QUILC{} performs the tasks of selecting those compilation subroutines suited to the problem at hand.  To demonstrate the flexibility afforded by this approach, we consider a simple experiment in compiling \texttt{CCNOT}, and in particular consider the effect that the choice of native gateset and availability of particular compilation subroutines has on the resulting gate complexity.

Recalling our previous example, we note that the circuit expressed in \Cref{fig:connected_ccnot} is embodied in \QUILC{} as a compilation subroutine, \texttt{CCNOT-to-CNOT} (cf.\ \Cref{fig:ccnot_to_cnot} and the discussion in \Cref{ImpDetailsSection}). Strictly speaking, specific subroutines such as this one are not needed, as \QUILC{} supports fully generic techniques such as the recursive ``Quantum Shannon Decomposition'' of \cite{shende2006synthesis}. Nonetheless, specific compilation subroutines may be preferred to general ones when they offer a reduction in final native gate counts. 

In what follows, we consider a three qubit chip with linear connectivity, so that qubit~0 is connected to 1, and qubit~1 is connected to 2. Amongst the possible two-qubit operations, we restrict attention to \texttt{CZ}, \texttt{ISWAP}, and \texttt{CPHASE}, and for each subset of these we consider a target architecture in which those operations are native across connected qubits. With respect to compilation subroutines, \QUILC{} has several enabled by default, and we consider only the effect of including or excluding \texttt{CCNOT-to-CNOT} from this set. In all instances, the compiler is able to translate \texttt{CNOT} gates to the native gate of choice and is able to take advantage of ``\texttt{SWAP} recombination'' as described earlier.
 
 In \Cref{tbl:CCNOT_2q_depth} we show the complexity of \texttt{CCNOT} in terms of native two-qubit gate counts. In all instances, it is always advantageous to incorporate the special information provided by \texttt{CCNOT-to-CNOT}. The best results occur for a device supporting \texttt{CZ} along with one of $\{ \texttt{ISWAP},\texttt{CPHASE} \}$, which results in a circuit using only 6 two-qubit gates.

\begin{table}[h]
\centering
\begin{tabular}{cccrr}
\toprule
\multicolumn{3}{c}{Native Gates} & \multicolumn{2}{c}{Gate Count} \\
\cmidrule(r){1-3} \cmidrule(r){4-5}
\texttt{CZ}  & \texttt{ISWAP} & \texttt{CPHASE} &  Without & With   \\
\midrule
\checkmark &  & & 9 & 7 \\
& \checkmark & & 12 & 9  \\
& & \checkmark & 8  & 8  \\
\checkmark & \checkmark & & 7 & 6  \\
\checkmark & & \checkmark & 8 & 6  \\
& \checkmark & \checkmark & 10 & 9  \\
\checkmark & \checkmark & \checkmark & 7 & 6  \\
\bottomrule
\end{tabular} %
\caption{Two-qubit native gate complexity of \texttt{CCNOT} with linear nearest-neighbor connectivity. Gate counts are shown both with and without the use of the \texttt{CCNOT-to-CNOT} compilation subroutine. Note the uniform improvement in gate count introduced by the addition of the subroutine, whether or not \texttt{CNOT} appears in the target instruction set.}
\label{tbl:CCNOT_2q_depth}
\end{table}

\subsection*{State-Aware Compilation}

Many compilation routines express simple circuit equivalencies, which depend only on the syntactic structure of some instruction or block of instructions. However, \QUILC{} is capable of providing more information to those routines which might benefit from this context. When \textit{state-aware} compilation is enabled, \QUILC{} performs a partial simulation of addressed \QUIL{} instructions during the compression stage, until such a simulation is obstructed, e.g. up to an entanglement limit for performance reasons (by default, interactions of up to three qubits), or until a run-time data dependency is encountered, due to the unavailability of such data at the time of compilation. The results of this simulation are made available to additional compilation routines. Those routines which make use of this additional information are called \emph{state-aware}.

As an example, note that in general a compilation subroutine translates a gate application or sequence of gate applications to some ``equivalent'' sequence. Under most circumstances, the notion of equivalence here is that the corresponding unitary transformations, represented by the instructions, should be equal, perhaps up to some discretization error. However, when the quantum state is known prior to the execution of an instruction or block of instructions, the requirement of unitary equivalence may be relaxed. Indeed, given full information about the initial state, a sufficient notion of equivalence is that the resulting states be equal. The corresponding task, of preparing a target state given an initial state, is known as {\it state preparation}. Many instruction sequences which are not unitarily equivalent may be equivalent in this sense, and this increase in flexibility allows for additional synthesis techniques~\cite{plesch2011prep},~\cite[Section 4]{shende2006synthesis}.

\QUILC{} incorporates a number of methods for state preparation, ranging from special purpose compilation subroutines (for example, in the case of one-qubit, two-qubit, or four-qubit systems), to generic subroutines which may recurse to one of these special cases. State preparation is fully compatible with the optimizations available through other compilation subroutines, although it is only applicable in circumstances in which the initial state may be effectively computed.

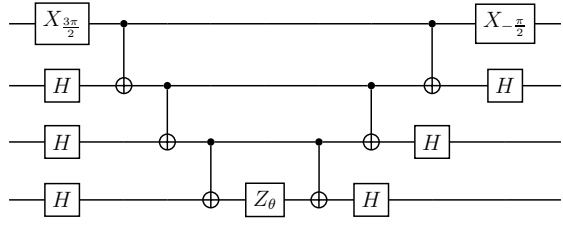
\begin{figure}
\centering
\adjustbox{scale=0.7,center}{%
\begin{tikzcd}
\qw & \gate{X_{\frac{3\pi}{2}}} & \ctrl{1} & \qw & \qw & \qw & \qw & \qw & \ctrl{1} & \gate{X_{-\frac{\pi}{2}}} & \qw \\
\qw & \gate{H} & \targ{} & \ctrl{1} & \qw & \qw & \qw & \ctrl{1} & \targ{} & \gate{H} & \qw \\
\qw & \gate{H} & \qw & \targ{} & \ctrl{1} & \qw & \ctrl{1} & \targ{} & \gate{H} & \qw & \qw \\
\qw & \gate{H} & \qw & \qw & \targ{} & \gate{Z_{\theta}} & \targ{} & \gate{H} & \qw & \qw & \qw
\end{tikzcd}
}
\caption{UCCSD ansatz for $\mathrm{H}_2$ in the STO-3G basis.}\label{fig:h2ansatz}
\end{figure}

We demonstrate this by way of an example. In \Cref{fig:h2ansatz}, we have a short program which expresses the unitary coupled cluster ansatz for deuterium, truncated to single and double excitation levels~\cite[Section VII.1]{mcardle2018quantum}. This program may be used as part of a hybrid classical/quantum variational algorithm such as described in \cite{peruzzo2014variational}. In such hybrid algorithms, it is typical that a single parametric program ``template'' is executed for a variety of numerical parameter values.\footnote{\QUILC{} supports {\it parametric compilation}, in which a program may be compiled once and then executed with parameter values determined at runtime. This method of compilation can be used to greatly improve hybrid computation~\cite{karalekas2020}.}

\begin{figure}
\centering 
\adjustbox{scale=0.7,center}{%
\begin{tikzcd}
\qw & \gate{Z_{\pi} X_{\frac{\pi}{2}}} & \qw & \qw & \qw & \qw & \gate{Z} & \gate{Z_{\pi} X_{-\frac{\pi}{2}}} & \qw \\
\qw & \gate{Z_{\frac{\pi}{2}} X_{\frac{\pi}{2}}} & \qw & \qw & \gate{Z} & \gate{X_{\frac{\pi}{2}}} & \ctrl{-1} & \gate{Z_{\frac{\pi}{2}} X_{\pi}} & \qw \\
\qw & \gate{Z_{\frac{\pi}{2}} X_{\frac{\pi}{2}} Z_{\frac{\pi}{2}}} & \gate{Z} & \gate{Z_{\frac{\pi}{2}} X_{\frac{\pi}{2}}} & \ctrl{-1} & \gate{Z_{\frac{\pi}{2}} X_{\pi}} & \qw & \qw & \qw \\
\qw & \gate{X_{\pi} Z_{-\frac{\pi}{2}} X_{-\frac{\pi}{2}} Z_{\theta} X_{\frac{\pi}{2}}} & \ctrl{-1} & \gate{Z_{\frac{\pi}{2}} X_{\pi}} & \qw & \qw & \qw & \qw & \qw
\end{tikzcd}
}
\caption{UCCSD ansatz for $\mathrm{H}_2$ in the STO-3G basis, compiled to native gates with state-aware optimizations enabled, requiring only 3 \texttt{CZ} instructions.} \label{fig:h2ansatzstateprep}
\end{figure}
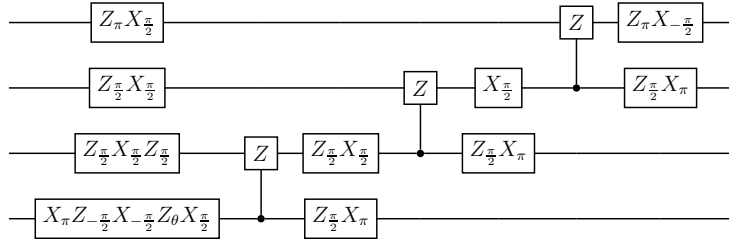

On near term devices, numerical accuracy of variational methods reflects a trade-off between the sophistication of the ansatz and the resulting depth or complexity of the circuit. When compiled to native hardware without state preparation routines, the resulting program involves 6 two-qubit operations as in \Cref{fig:h2ansatznaive}. When state preparation routines are allowed, knowledge that the program begins in the zero state $|0000\rangle$ is exploited to reduce this number to 3, as in \Cref{fig:h2ansatzstateprep}. 

In this example, one can see the practical effect of state-aware compilation is to reduce the initial portion of the circuit up until a point at which the state is no longer feasible to track. Here we remark that the presence of run-time parameters obstructs the partial-state simulation, and hence the reduction in gate count is primarily due to optimizations in the first half of the circuit. Additional details of state-aware compilation are discussed in Section ~\ref{ImpDetailsSection}.

\begin{figure}
\centering
\adjustbox{scale=0.7,center}{%
\begin{tikzcd}
\qw & \gate{X_{-\frac{\pi}{2}}} & \gate{Z} & \gate{Z_{\pi}} & \qw & \qw & \qw & \qw & \qw & \qw & \qw & \qw & \gate{Z} & \gate{Z_{\pi} X_{-\frac{\pi}{2}}} & \qw \\
\qw & \qw & \ctrl{-1} & \gate{Z_{-\frac{\pi}{2}} X_{-\frac{\pi}{2}}} & \gate{Z} & \qw & \qw & \qw & \qw & \qw & \gate{Z} & \gate{X_{\frac{\pi}{2}}} & \ctrl{-1} & \gate{Z_{\frac{\pi}{2}} X_{\pi}} & \qw \\
\qw & \gate{X_{\pi}} & \qw & \qw & \ctrl{-1} & \gate{Z_{-\frac{\pi}{2}} X_{-\frac{\pi}{2}}} & \gate{Z} & \gate{Z_{-\frac{\pi}{2}}} & \gate{Z} & \gate{Z_{\frac{\pi}{2}} X_{\frac{\pi}{2}}} & \ctrl{-1} & \gate{Z_{\frac{\pi}{2}} X_{\pi}} & \qw & \qw & \qw \\
\qw & \gate{X_{\pi}} & \qw & \qw & \qw & \qw & \ctrl{-1} & \gate{Z_{-\frac{\pi}{2}} X_{-\frac{\pi}{2}} Z_{\theta} X_{\frac{\pi}{2}}} & \ctrl{-1} & \gate{Z_{\frac{\pi}{2}} X_{\pi}} & \qw & \qw & \qw & \qw & \qw
\end{tikzcd}
}
\caption{UCCSD ansatz for $\mathrm{H}_2$ in the STO-3G basis, naively compiled to Rigetti's native gates, requiring 6 \texttt{CZ} instructions.} \label{fig:h2ansatznaive}
\end{figure}
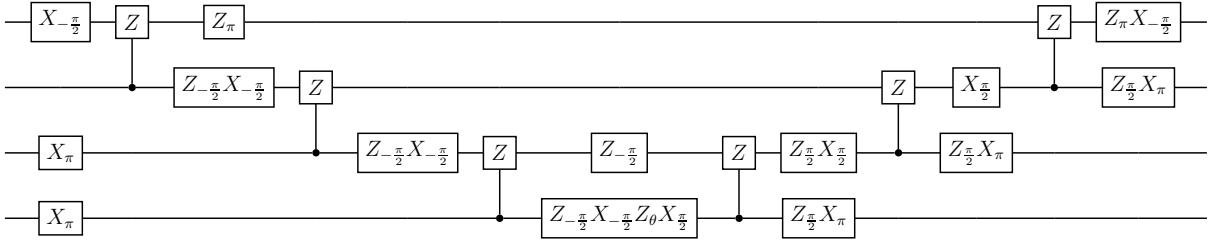

\section{Performance}\label{sec:Performance}

There are two meanings to the word ``performance'' when it comes to compilers: its effectiveness at compiling a quantum program, and how many resources it consumes to perform that task.

To measure compilation effectiveness, we use the benchmarks from \cite{zulehner2019efficient}---a suite of QASM files---that test \QUILC{} holistically in two ways: with state-aware compilation disabled (i.e., the input and output of the compiler are equivalent unitaries), and with state-aware compilation enabled (i.e., the input and output of the compiler are only guaranteed to act identically on the ground state). \Cref{tab:qasmbench} contains the benchmarks.

To date, \QUILC{} has not been optimized for time or space performance (i.e., how long it takes to compile and how much memory is required), and with an appropriate combination of engineering tricks, those metrics could be improved.  Even in the absence of these optimizations, the run-time statistics presented in \Cref{QFTBenchmarkFigure} and \Cref{BenchmarkFigure} demonstrate the scaling laws associated with our approximate solvers deployed during the addressing and compressing stages. Wall clock performance is also included in \Cref{tab:qasmbench}. 

{\tiny
\begin{longtable}{lrrrrr}
\caption{QASM benchmarks from~\cite{zulehner2019efficient} targeting the IBM qx5 architecture, performed in the same environment as \Cref{BenchmarkFigure}. Compare with \cite[Table I]{zulehner2019efficient}. Columns labeled with an asterisk `*' hold data for circuits which were produced with the additional hypothesis that the quantum device begins in the ground state. The percentage in parentheses column indicates the rounded percentage reduction in 2Q \emph{depth} by taking this assumption into account.}\label{tab:qasmbench}\\
\hline
\textbf{Filename} & \textbf{\texttt{SWAP}s Added} & \textbf{Wall Time (s)} & \textbf{2Q Depth} & \textbf{Wall Time* (s)} & \textbf{2Q Depth* (\%)}\\
\hline
\endfirsthead
\multicolumn{6}{c}%
{\tablename\ \thetable\ -- \textit{Continued from previous page}} \\
\hline
\textbf{Filename} & \textbf{\texttt{SWAP}s Added} & \textbf{Wall Time (s)} & \textbf{2Q Depth} & \textbf{Wall Time* (s)}& \textbf{2Q Depth* (\%)}\\
\hline
\endhead
\hline \multicolumn{6}{r}{\textit{Continued on next page}} \\
\endfoot
\hline
\endlastfoot
0410184\_169          &   101 &    8.767 &      235 &    9.802 &       55 	 (77\%)\\
3\_17\_13             &    11 &    0.485 &       28 &     0.79 &       23 	 (18\%)\\
4\_49\_16             &    66 &    7.271 &      230 &   10.738 &      109 	 (53\%)\\
4gt10-v1\_81          &    44 &    5.311 &      143 &    8.771 &       86 	 (40\%)\\
4gt11\_82             &     5 &    1.705 &       28 &    1.387 &        7 	 (75\%)\\
4gt11\_83             &     4 &    0.809 &       22 &    0.837 &        0 	 (100\%)\\
4gt11\_84             &     3 &    0.436 &       15 &    0.544 &       13 	 (13\%)\\
4gt12-v0\_86          &    76 &    9.364 &      241 &    13.11 &      117 	 (51\%)\\
4gt12-v0\_87          &    73 &   10.018 &      231 &    12.33 &      116 	 (50\%)\\
4gt12-v0\_88          &    52 &     6.11 &      176 &    8.015 &       87 	 (51\%)\\
4gt12-v1\_89          &    67 &    7.353 &      234 &    9.621 &       91 	 (61\%)\\
4gt13-v1\_93          &    16 &    1.878 &       60 &    3.257 &       31 	 (48\%)\\
4gt13\_90             &    28 &    3.111 &      102 &    5.355 &       47 	 (54\%)\\
4gt13\_91             &    27 &     2.92 &       97 &    4.105 &       30 	 (69\%)\\
4gt13\_92             &    19 &    1.195 &       57 &    1.913 &       23 	 (60\%)\\
4gt4-v0\_72           &    66 &    7.307 &      218 &   12.685 &       88 	 (60\%)\\
4gt4-v0\_73           &   108 &   12.503 &      371 &   20.436 &      183 	 (51\%)\\
4gt4-v0\_78           &    67 &    7.194 &      233 &   10.774 &       77 	 (67\%)\\
4gt4-v0\_79           &    66 &     6.69 &      225 &    10.17 &       91 	 (60\%)\\
4gt4-v0\_80           &    57 &    5.896 &      185 &    9.777 &       87 	 (53\%)\\
4gt4-v1\_74           &    78 &    7.352 &      260 &   15.558 &      184 	 (29\%)\\
4gt5\_75              &    24 &    3.108 &       88 &    4.995 &       38 	 (57\%)\\
4gt5\_76              &    27 &    3.235 &       83 &    6.879 &       45 	 (46\%)\\
4gt5\_77              &    37 &    3.547 &      119 &    7.457 &       84 	 (29\%)\\
4mod5-bdd\_287        &    19 &     1.75 &       75 &     3.04 &       25 	 (67\%)\\
4mod5-v0\_18          &    20 &     1.23 &       69 &    1.781 &       46 	 (33\%)\\
4mod5-v0\_19          &     9 &    0.694 &       35 &    1.103 &       21 	 (40\%)\\
4mod5-v0\_20          &     6 &    0.875 &       21 &    1.956 &       14 	 (33\%)\\
4mod5-v1\_22          &     4 &    0.354 &       18 &    0.512 &       14 	 (22\%)\\
4mod5-v1\_23          &    19 &    2.024 &       68 &    2.855 &       37 	 (46\%)\\
4mod5-v1\_24          &     8 &    1.474 &       32 &    2.435 &       19 	 (41\%)\\
4mod7-v0\_94          &    51 &    4.291 &      155 &    6.248 &       89 	 (43\%)\\
4mod7-v1\_96          &    50 &     5.16 &      172 &    6.595 &       62 	 (64\%)\\
C17\_204              &   138 &   14.616 &      436 &    20.53 &      268 	 (39\%)\\
aj-e11\_165           &    49 &    5.232 &      144 &    7.957 &       59 	 (59\%)\\
alu-bdd\_288          &    22 &    1.382 &       78 &    1.721 &       39 	 (50\%)\\
alu-v0\_26            &    22 &    3.405 &       88 &    6.048 &       51 	 (42\%)\\
alu-v0\_27            &     8 &    1.239 &       30 &    2.459 &       15 	 (50\%)\\
alu-v1\_28            &     8 &    1.083 &       33 &    1.765 &       14 	 (58\%)\\
alu-v1\_29            &     7 &    1.098 &       33 &    1.683 &       17 	 (48\%)\\
alu-v2\_30            &   140 &   14.819 &      471 &   22.585 &      228 	 (52\%)\\
alu-v2\_31            &   146 &   15.981 &      428 &    21.86 &      182 	 (57\%)\\
alu-v2\_32            &    57 &    6.641 &      160 &   11.628 &       82 	 (49\%)\\
alu-v2\_33            &     8 &    0.965 &       28 &    1.637 &       16 	 (43\%)\\
alu-v3\_34            &    12 &    1.629 &       50 &     2.34 &       22 	 (56\%)\\
alu-v3\_35            &    11 &      1.1 &       36 &     1.86 &       28 	 (22\%)\\
alu-v4\_36            &    29 &    2.242 &       97 &    3.456 &       53 	 (45\%)\\
alu-v4\_37            &     8 &    1.247 &       31 &    2.318 &       16 	 (48\%)\\
cnt3-5\_179           &    63 &    6.172 &      139 &    6.467 &       45 	 (68\%)\\
cnt3-5\_180           &   139 &    9.593 &      385 &   16.345 &      183 	 (52\%)\\
decod24-bdd\_294      &    22 &    1.497 &       73 &    2.423 &       51 	 (30\%)\\
decod24-enable\_126   &   122 &   11.186 &      364 &   18.681 &      158 	 (57\%)\\
decod24-v0\_38        &    10 &    0.745 &       43 &    5.446 &       31 	 (28\%)\\
decod24-v1\_41        &    25 &    1.319 &       76 &    2.143 &       34 	 (55\%)\\
decod24-v2\_43        &    15 &    1.181 &       53 &    1.686 &       10 	 (81\%)\\
decod24-v3\_45        &    41 &    2.715 &      130 &     4.79 &       70 	 (46\%)\\
ex-1\_166             &     6 &    0.314 &       15 &    0.563 &       11 	 (27\%)\\
ex1\_226              &     1 &    0.564 &        6 &    0.786 &        0 	 (100\%)\\
ex2\_227              &   179 &   21.328 &      612 &   27.749 &      268 	 (56\%)\\
ex3\_229              &   100 &    9.691 &      367 &   18.069 &      223 	 (39\%)\\
graycode6\_47         &     0 &    0.638 &        5 &    0.774 &        0 	 (100\%)\\
ham3\_102             &     6 &    0.355 &       15 &    0.454 &       13 	 (13\%)\\
ham7\_104             &    72 &    9.683 &      307 &   15.797 &      149 	 (51\%)\\
hwb4\_49              &    74 &    6.803 &      227 &   11.394 &      119 	 (48\%)\\
ising\_model\_10      &     0 &    1.665 &       20 &     2.31 &       19 	 (5\%)\\
ising\_model\_13      &     0 &    2.862 &       20 &    4.738 &       19 	 (5\%)\\
ising\_model\_16      &     0 &    3.748 &       20 &    6.182 &       19 	 (5\%)\\
miller\_11            &    12 &     0.55 &       47 &    0.384 &        0 	 (100\%)\\
mini-alu\_167         &    94 &    9.481 &      272 &   16.332 &      125 	 (54\%)\\
mini\_alu\_305        &    62 &    4.388 &      173 &    7.373 &       39 	 (77\%)\\
mod10\_171            &    69 &    4.293 &      216 &    6.471 &       78 	 (64\%)\\
mod10\_176            &    41 &    5.314 &      164 &    8.637 &       81 	 (51\%)\\
mod5adder\_127        &   164 &   15.054 &      498 &   24.652 &      286 	 (43\%)\\
mod5d1\_63            &     6 &    0.591 &       25 &    0.804 &       11 	 (56\%)\\
mod5d2\_64            &    21 &     2.52 &       70 &    3.943 &       34 	 (51\%)\\
mod5mils\_65          &    10 &    1.489 &       34 &    2.825 &       20 	 (41\%)\\
mod8-10\_177          &   127 &   13.593 &      412 &   22.071 &      195 	 (53\%)\\
mod8-10\_178          &    85 &    8.877 &      310 &   13.051 &      146 	 (53\%)\\
one-two-three-v0\_97  &    90 &    6.717 &      263 &   12.593 &      167 	 (37\%)\\
one-two-three-v0\_98  &    48 &    4.321 &      148 &    7.224 &       73 	 (51\%)\\
one-two-three-v1\_99  &    36 &     4.06 &      128 &    6.711 &       61 	 (52\%)\\
one-two-three-v2\_100 &    21 &    2.998 &       69 &     4.28 &       26 	 (62\%)\\
one-two-three-v3\_101 &    19 &    1.656 &       72 &    2.529 &       36 	 (50\%)\\
qft\_10               &    28 &    2.805 &       44 &    3.491 &        0 	 (100\%)\\
qft\_16               &    86 &    7.495 &      111 &    9.063 &        6 	 (95\%)\\
rd32-v0\_66           &     8 &    0.495 &       26 &    0.877 &       21 	 (19\%)\\
rd32-v1\_68           &     8 &    0.501 &       26 &    0.865 &       21 	 (19\%)\\
rd32\_270             &    20 &    2.985 &       81 &    4.421 &       31 	 (62\%)\\
rd53\_131             &   130 &   15.426 &      439 &   24.783 &      186 	 (58\%)\\
rd53\_135             &    94 &   10.164 &      312 &   17.044 &      178 	 (43\%)\\
rd53\_138             &    38 &     4.54 &      122 &    5.982 &       48 	 (61\%)\\
rd53\_311             &    98 &    9.199 &      258 &   14.522 &      119 	 (54\%)\\
rd73\_140             &    68 &    7.166 &      211 &   10.845 &      104 	 (51\%)\\
rd84\_142             &   110 &    7.454 &      227 &    9.077 &      110 	 (52\%)\\
sf\_274               &   204 &   16.251 &      691 &   28.094 &      458 	 (34\%)\\
sym6\_316             &    98 &    7.566 &      269 &   14.242 &       99 	 (63\%)\\
sym9\_146             &    91 &     8.76 &      257 &   15.803 &      126 	 (51\%)\\
sys6-v0\_111          &    67 &    6.076 &      158 &    9.943 &       60	 (62\%)\\
\end{longtable}
}

\begin{figure}[!h]\label{fig:perf}\caption{Run-time performance measurements of \QUILC{}.}
\begin{subfigure}{\textwidth}
\begin{tikzpicture}
\begin{groupplot}
[
    group style={
        group name=my plots,
        group size=2 by 1,
        xlabels at=edge bottom,
        xticklabels at=edge bottom,
        vertical sep=1cm
    },
    footnotesize,
    width=8cm,
    height=5cm,
    xlabel=Qubits,
    xmin=0,
    ymin=0,
    tickpos=left,
    ytick align=outside,
    xtick align=outside,
]
\nextgroupplot[xlabel=Qubits vs.\ Run time (s)]
\addplot table[y=user_run_time] {QFT-program-compilation-fully-connected-nQ-chip.csv};
\addplot table[y=user_run_time] {QFT-program-compilation-Aspen-4-16Q.csv};
\nextgroupplot[xlabel=Qubits vs.\ Memory usage (MB), ymin=105, max space between ticks=1000, try min ticks=8]
\addplot table[y=megabytes_required] {QFT-memory-floor-fully-connected.csv};
\addplot table[y=megabytes_required] {QFT-memory-floor-32Q.csv};
\end{groupplot}
\end{tikzpicture}
    \centering
    \caption{Performance characteristics for compilation of QFT circuits of varying sizes to an Aspen-type device~(\textcolor{red}{red squares}) and a fully connected device~(\textcolor{blue}{blue circles}).  The left plot depicts the runtime of \QUILC{}, and the right plot depicts the minimum required heap size of \QUILC{} (to within 2 MB). \QUILC{} was compiled with Steel Bank Common Lisp (SBCL) version 1.5.8 and run on a 2018 MacBook Pro.}
    \label{QFTBenchmarkFigure}
\end{subfigure}
\newline
\begin{subfigure}{\textwidth}
\begin{tikzpicture}
\begin{groupplot}
[
    group style={
        group name=my other plots,
        group size=2 by 1,
        xlabels at=edge bottom,
        xticklabels at=edge bottom,
        vertical sep=1cm
    },
    footnotesize,
    width=8cm,
    height=5cm,
    xlabel=Qubits,
    xmin=0,
    ymin=0,
    ymax=400,
    tickpos=left,
    ytick align=outside,
    xtick align=outside,
]
\nextgroupplot[xlabel=Problem size vs.\ Run time (s),max space between ticks=1000, try min ticks=8]
\addplot[only marks] table[y=wall_time_secs] {qaoa-commuting.csv};
\nextgroupplot[xlabel=Problem size vs.\ Run time (s),max space between ticks=1000, try min ticks=8]
\addplot[only marks] table[y=wall_time_secs] {qaoa-noncommuting.csv};
\end{groupplot}
\end{tikzpicture}
    \centering
    \caption{Performance characteristics for compilation of random 3-valent QAOA-type problems of varying sizes to an 128-qubit Aspen-type device, with and without commutation information (left and right respectively).  Both plots depict the run-time of \QUILC{}, and both are best fit by a quadratic with small leading coefficient. \QUILC{} was compiled with Steel Bank Common Lisp (SBCL) version 1.5.5 and run on a 2016 MacBook Pro. See \Cref{listingqaoa} for an example of a QAOA-type program.}
    \label{BenchmarkFigure}
\end{subfigure}
\end{figure}

\section{Contributions \& Acknowledgements}

\QUILC{} is distributed as free and open source software under a permissive license. Contributions from anybody are highly welcome. Contributions come in a variety of forms, including bug reports, feature suggestions, UI/UX criticisms, platform testing, new circuit identities, or new compilation algorithms. The source code and issue tracker can be found here:
\begin{center}
\url{https://github.com/rigetti/quilc}
\end{center}

The authors would like to thank Rigetti Computing for its vibrant and stimulating workplace and Quantum Science and Technology (and B.\ Peropadre in particular) for the invitation to contribute to their special issue.  The authors are also grateful for the helpful comments of those readers of an early draft: M.\ Appleby, J.\ Bello-Rivas, M.\ Brett, and M.\ Hodson, as well as the anonymous referees.  The authors are all themselves significant contributors to \QUILC{}, but they have also all benefited from others' contributions to \QUILC{} as an open-source project, which have bolstered the project generally and shaped some of the ideas presented in this paper. Further contributor acknowledgements can be found in \QUILC{}'s \href{https://github.com/rigetti/quilc/blob/master/ACKNOWLEDGEMENTS.md}{\texttt{ACKNOWLEDGEMENTS.md}} file.  All authors contributed equally to this paper.


\bibliographystyle{plain}
\bibliography{references}

\appendix
\label{appendix}

\section{Implementation Details}
\label{ImpDetailsSection}

\subsection{Implementation Language}\label{sec:lisp}

\QUILC{} is written in ANSI Common Lisp, and can be extended with code written in languages with C ABI compatibility. Common Lisp was chosen because it provides a highly-performant substrate for both dynamic, interactive work as well as batch-mode computation, and offers convenient abstractions for implementing embedded domain-specific languages (DSLs), which are useful for many of the tasks of program analysis and manipulation. \QUILC{} is also compatible with Rigetti's open-source quantum computer simulator, the Quantum Virtual Machine~\cite{QVM}.

One reason Common Lisp is particularly suitable for implementing DSLs is its syntax. Common Lisp has very regular syntax, most of which follows just a single syntactic construction:
\begin{displaymath}
\texttt{(}\langle\textit{operator}\,\rangle\;\langle\textit{operand}\,\rangle_1\;\ldots\;\langle\textit{operand}\,\rangle_n\texttt{)}.
\end{displaymath}
That is, operators precede their operands and are surrounded by a pair of parentheses; see \Cref{tbl:lispsyntax} for a sample of syntax. Parentheses don't indicate precedence, but instead play double duty: as prefix-notation for the language, but also as syntax for (sometimes nested) lists. Since the syntax can be viewed as both notation and a data structure, Lisp code is ripe for both automatic generation and manipulation. For this reason, we call Lisp a \emph{homoiconic language}. Interested readers can find lengthy discussions of these ideas at different levels in the literature~\cite{sicp,norvig}. In the subsections to follow, excerpts from \QUILC{}'s source code will make use of this syntax.
\begin{table}[h!]
\centering
\begin{tabular}{lp{5cm}}
     Algebraic Syntax & Lisp's Syntax \\
     \hline
     $2(1-n)$        & \verb|(* 2 (- 1 n))|\\
     $f(x,g(y),x+y)$ & \verb|(f x (g y) (+ x y))|\\
     $k \mapsto k^3$ (i.e., $\lambda k.k^3$) & \verb|(lambda (k) (expt k 3))|\\
     let $z=\sqrt{5}$ in $\zeta(z/2)$ & \texttt{(let ((z (sqrt 5)))}\newline\verb|  |\texttt{(zeta (/ z 2)))}
\end{tabular}
\caption{A comparison between usual algebraic syntax as found in mathematics and many programming languages and the syntax of Common Lisp.}
\label{tbl:lispsyntax}
\end{table}

The \QUILC{} source code is separated into an application domain and a library domain.  The application can be used to consume textual input as a UNIX command-line tool, or it can be used to provide a persistent server front-end.  The library domain includes all of the routines for interpreting and manipulating \QUIL{} code, as well as some features that benefit from that availability but which do not directly participate in the compilation pipeline (e.g., generation and manipulation of Clifford group elements).

\subsection{Compiler Subroutine DSL}\label{sec:defcomp}

\QUILC{} implements a domain-specific language (DSL) for writing compilation subroutines. This makes it easy to specify algebraic relationships which \QUILC{} can use as a part of its automatic process of program decomposition and optimization.

The basic method of definition is \texttt{define-compiler}, whose basic syntax is given by the following \texttt{defun} mimic:
\begin{lstlisting}
(define-compiler name ([binding-with-options ...]
                       [global-option ...])
  [body-form ...])
\end{lstlisting}
This defines a compiler which consumes an instruction argument for each binding, evaluates the body forms in order, and returns the collection of instructions that they aggregate.  Each binding is specified by a variable in which the input gate is stored, as well as an optional destructuring pattern to capture its operator name, parameter list, and argument list.  This can be further manipulated by specifying options, which might install a further matching predicate on the destructured information.  Altogether, these take the following form:
\begin{lstlisting}
binding-with-options := (name (operator parameter-list argument-list)
                         [local-option ...])
                      | (name [local-option ...])
                      | name
                      
parameter-list := ([parameter ...])
argument-list  := [argument ...]
local-option   := :where PREDICATE | ...

operator  := SYMBOL | WILDCARD | STRING
parameter := SYMBOL | DECIMAL-LITERAL
argument  := SYMBOL | INDEX-LITERAL
\end{lstlisting}
The forms describing the body of the compiler are largely identical to forms elsewhere in Lisp, but there are a few special-use forms available as well.  The \texttt{inst} and \texttt{inst*} operators send an instruction to the output queue, which gets emptied for use as the return value at the conclusion of the compiler body.  Alternatively, \texttt{finish-compiler} and \texttt{give-up-compilation} can be used to manually signal the end of the compiler body, additionally providing optional manual control over what is used for the return value.
\begin{lstlisting}
body-form      := (inst INSTRUCTION)
                | (inst operator parameter-list argument-list)
                | (inst operator matrix argument-list)
                | (inst* operator parameter-list argument-list
                                                (argument-list)) 
                | (inst* operator matrix argument-list
                                        (argument-list))
                | (finish-compiler [return-form])
                | (give-up-compilation)
                | FORM
\end{lstlisting}

We now demonstrate the construction of compilation routines of increasing levels of complexity. We remark that these examples are taken to be didactic; \QUILC{} houses roughly a hundred distinct compilation routines, ranging from algebraic rewriting rules specialized on gates which are common targets for implementation on current and near term hardware (e.g. \texttt{RZ}, \texttt{CNOT}, \texttt{CZ}) to general purpose recursive routines applicable to arbitrary unitary operations.

\begin{example}
Linearity of certain gates allows the compiler to collapse sequences of applications of a single gate. The optimizer \texttt{agglutinate-RZs} below rests on the linearity property \[\mathtt{RZ}(\theta) \cdot \mathtt{RZ}(\phi) = \mathtt{RZ}(\theta + \phi).\]  It matches against two \texttt{RZ} instructions (bound to the variables \texttt{x} and \texttt{y}) acting on a particular common qubit \texttt{q}, and it binds their parameter values to the variables \texttt{theta} and \texttt{phi} respectively. The compiler then emits a single instruction \texttt{RZ($\theta + \phi$)} that replaces the two input instructions.
\begin{lstlisting}[language=lisp]
(define-compiler agglutinate-RZs
    ((x ("RZ" (theta) q))
     (y ("RZ" (phi)   q)))
  (inst "RZ" `(,(param-+ theta phi)) q))
\end{lstlisting}
We note that on Rigetti's current hardware, the \texttt{RZ} gate is the primary target for parametric compilation, and the above routine is capable of acting both on numeric and symbolic values of $\theta$ and $\phi$ in a manner in which numeric routines (e.g. depending on the underlying gate matrix) would not be.
\end{example}

\begin{example}
Parametric gates for some parameter values will be equivalent to the identity operation (\texttt{NOP}). This is easily seen for those gates that impart a rotation upon a qubit's state about some axis: a rotation of $2\pi$ is equivalent to not having rotated at all.\footnote{The compiler here takes the idealist's view: that rotations are perfect and without noise. If we were to consider noise, then it might not be the case that \texttt{RX($2\pi$)} is the identity.} The \texttt{eliminate-full-CPHASE} optimizer implements this reduction for the instruction \texttt{CPHASE}. The compiler matches against a \texttt{CPHASE} instruction, binds the parameter value to \texttt{theta}, and via the \texttt{:where} guard restricts input instructions to those whose parameter values are integer multiples of $2\pi$. The optimizer emits no instructions, thereby reducing the instruction stream by one.
\begin{lstlisting}[language=lisp]
(define-compiler eliminate-full-CPHASE
    ((x ("CPHASE" (theta) _ _)
        :where (and (typep theta 'double-float)
                    (double= 0d0 (mod theta (* 2 pi))))))
  nil)
\end{lstlisting}
\end{example}

\begin{example}
If the compiler can find no compilation routines that specifically match against a given input instruction, then it will look for less specific routines. The nativizer \texttt{euler-zyz-compiler} below only requires that it match against a single-qubit instruction, without specifying the name of the instruction or its parameters.\footnote{Note that in this routine, explicit numeric values are expected for the gate entries in order to compute the CSD composition. With respect to parametric compilation, such routines are inapplicable at sites where symbolic parameters are present.} Since any single-qubit instruction is equivalent to a rotation about some axis, it can be decomposed into three rotations about the $Y$ and $Z$ axes. This compilation routine is unlike the previous examples in that it emits more instructions than it consumes.\footnote{This subroutine makes use of \texttt{magicl}, a Common Lisp library for numerical linear algebra.}
\begin{lstlisting}[language=lisp,mathescape]
(define-compiler euler-zyz-compiler ((instr (_ _ q)))
  ;; Cosine-sine decomposition (CSD) of a unitary matrix $U$ is
  ;; described as $U = (u_0 \oplus u_1)D(v_0 \oplus v_1)$ with
  ;;   * $u_0$, $u_1$, $v_0$, $v_1$ unitary;
  ;;   * $D_{00} = D_{11} = \operatorname{diag}\{\cos \theta : \theta\in\mathit{angles}\}$; and
  ;;   * $D_{10} = -D_{01} = \operatorname{diag}\{\sin \theta : \theta\in\mathit{angles}\}$.
  (multiple-value-bind (u0 u1 v0 v1 angles)
      (magicl:lapack-csd (gate-matrix instr) 1 1)
    (let ((alpha (- (phase (magicl:ref v1 0 0))     ; (magicl:ref $M$ $i$ $j$) = $M_{i,j}$
                    (phase (magicl:ref v0 0 0))))
          (beta (first angles))
          (gamma (- (phase (magicl:ref u1 0 0))
                    (phase (magicl:ref u0 0 0)))))
     (inst "RZ" (list alpha) q)
     (inst "RY" (list beta)  q)
     (inst "RZ" (list gamma) q)))))
\end{lstlisting}
\end{example}

\begin{example}
Given the preceding routines, we give an example trace of the compiler for a simple \QUIL{} program. The ordering of steps below is \emph{automatically} determined by \QUILC{}, and changes with target architecture and other factors. In particular, with respect to the taxonomy of \Cref{sec:structure}, \texttt{agglutinate-RZs} and \texttt{eliminate-full-CPHASE} are \emph{optimizing} compilation routines, whereas \texttt{euler-zyz-compiler} is a \emph{nativizing} routine.

In this sample trace, we consider a hypothetical two-qubit device which supports \texttt{RZ}, \texttt{RY}, and \texttt{CPHASE} natively, but for which \texttt{H} is not a native gate. We start with the following program:
\begin{lstlisting}
    RZ(-pi)      0
    CPHASE(2*pi) 1 0
    H            0
\end{lstlisting}
During the addressing phase, the non-native Hadamard gate will be translated to native gates by means of \texttt{euler-zyz-compiler}, yielding:
\begin{lstlisting}
    RZ(-pi)      0
    CPHASE(2*pi) 1 0
    RZ(pi)       0
    RY(pi/2)     0
    RZ(0)        0
\end{lstlisting}
At this point, the ``logical'' qubits 0 and 1 have been assigned to ``physical'' qubits 0 and 1, without the need for any \texttt{SWAP} operations.
During the compression stage, \QUILC{} scans the program and will first apply \texttt{eliminate-full-CPHASE} to the spurious gate, yielding:
\begin{lstlisting}
    RZ(-pi)  0
    RZ(pi)   0
    RY(pi/2) 0
    RZ(0)    0
\end{lstlisting}
Finally, it applies \texttt{agglutinate-RZs} to the first pair of $Z$-rotations to produce:
\begin{lstlisting}
    RZ(0)    0
    RY(pi/2) 0
    RZ(0)    0
\end{lstlisting}
If we were to additionally tell \QUILC{} that $\mathtt{RZ}(0)$ is the identity:
\begin{lstlisting}[language=lisp]
(define-compiler ((instr ("RZ" (0d0) _)))
  nil)
\end{lstlisting}
then \QUILC{} will strike the outer instructions, leaving just:
\begin{lstlisting}
    RY(pi/2) 0
\end{lstlisting}

One can explore \QUILC{}'s treatment of an input program with the \verb|--verbose| option at the command line.
\end{example}

\subsection{Other Features of \QUILC{}}

We've described the backbone of \QUILC{}'s operation but have not described exhaustively all of \QUILC{}'s features. We mention a few additional salient features here.

\paragraph{Details of state-aware compilation} \QUILC{} includes a minimalistic \QUIL{} interpreter which is optionally used for various optimizations. If the user indicates that the state of the quantum system is initialized to $\ket{0}$, as is defined by \QUIL{}, then \QUILC{} will attempt to partially simulate the program up to a certain entanglement limit. The partially simulated state is then supplied to compilers defined with \verb|define-compiler|, where it can be optionally used. For instance, we can detect if the state is an eigenvector of an instruction and consequently eliminate it:
\begin{lstlisting}[language=lisp,mathescape]
(define-compiler elide-applications-on-eigenvectors
    ((instr :acting-on (psi qubit-indices)
            ;; (collinearp $u$ $v$) returns true iff $v=e^{i\theta}u$ for some $\theta$.
            :where (collinearp
                    psi
                    (nondestructively-apply-instr-to-wf instr psi qubit-indices))))
  nil)
\end{lstlisting}
More generally, if $U$ represents our original partial program and $\ket{\psi}=U\ket{0}$ represents the partially simulated state, then \QUILC{} can find an alternative $V$ such that $\ket{\psi}=V\ket{0}$ which compiles to fewer instructions than $U$ does. These features are enabled at the command line with \verb|--enable-state-prep-reductions|.

\paragraph{Permutation matrices and contributed modules} \QUILC{} does not have native support for compiling certain special kinds of unitary matrices, such as permutation matrices or diagonal matrices. This, in turn, makes \QUILC{} particularly poorly suited for compilation of classical reversible logic. However, due to its flexible mechanism for extension and automatic selection of compilers, and due to easy integration of libraries written in C or C++, \QUILC{} makes use of a library called Tweedledum~\cite{tweedledum}. Tweedledum has specialized routines for synthesizing such circuits and can be employed automatically by \QUILC{}.

\paragraph{Approximate compilation} \QUILC{} is able to profitably make use of gate fidelities to produce programs which have better overall fidelity. The fundamental observation is as follows. If a program $U = U_m\ldots U_1$ is expressed as $m$ native gates, then the program as executed on a quantum computer will be some other $U' = U'_m\ldots U'_1$, where the difference or uncertainty of $U_i$ and $U'_i$ is reflected in the native gates' fidelity. We might deliberately compile $U$ as some \emph{non-equivalent} sequence of native gates $V=V_n\ldots V_2V_1$ such that, when run on a quantum computer with initial state $\ket{\psi_0}$ and realized as $V'=V'_n\ldots V'_2V'_1$, the average fidelity is increased: \[\int |\langle \psi| U^\dagger U' | \psi \rangle|^2\,d\psi < \int |\langle \psi| U^\dagger V' | \psi \rangle|^2\,d\psi.\]  The existence of such alternative circuits is discussed in \cite{PhysRevA.100.032328, PCS}.

\paragraph{Parametric compilation} \QUILC{} is capable of dealing with symbolic parameters in a variety of cases beyond the modest use in \Cref{sec:examples}.  As a small example, \QUILC{} reduces the following snippet
\begin{lstlisting}
DECLARE a REAL
RZ(a) 0
RZ(0.5*a) 0
RZ(0.2) 0
\end{lstlisting}
to the single instruction \verb|RZ(0.2+1.5*a) 0|. As a more complicated example, the instruction \verb|CPHASE(t/3) 0 1| compiles to 
\begin{lstlisting}
RZ(-pi/2) 1
RX(pi/2) 1
CZ 1 0
RX(-pi/2) 1
RZ(-t/6) 1
RX(pi/2) 1
CZ 1 0
RZ(t/6) 0
RX(-pi/2) 1
RZ(pi/2 + t/6) 1
\end{lstlisting}
on a \verb|CZ|-based architecture.

\paragraph{Hamiltonians consisting of Pauli sums} Sums of Pauli operators occur frequently, especially in Hamiltonians used to describe qubit-qubit interactions. For instance, for qubits $a$ and $b$, the ``$\mathrm{XY}$'' interaction is defined by the Hamiltonian \[H_{\mathrm{XY}} := X_a\otimes X_b + Y_a\otimes Y_b,\] or simply $XX+YY$ when context is clear. The interaction itself is described by the unitary operation
\begin{displaymath}
U_{\mathrm{XY}}(\theta) := \exp(i\theta H_{\mathrm{XY}}).
\end{displaymath}
\QUIL{} is able to specify such unitary operators with the \texttt{DEFGATE} (see \Cref{listingqaoa}) construct:
\begin{lstlisting}
DEFGATE Uxy(%theta) a b AS PAULI-SUM:
    XX(%theta) a b
    YY(%theta) a b
\end{lstlisting}
\QUILC{} is able to compile operators defined in such a way, even in some cases for undetermined parameters via parametric compilation of the last section. For instance, consider the program:
\begin{lstlisting}
DECLARE t REAL
Uxy(t) 0 1
\end{lstlisting}
Providing this \QUILC{} will emit the following 12-gate program
\begin{lstlisting}
RZ(-pi/2)    0
RX(pi)       0
RX(pi/2)     1
CZ           0 1
RZ(pi)       0
RX(pi)       0
RX(pi/2)     1
RZ(pi - 2*t) 1
RX(pi/2)     1
CZ           0 1
RZ(-pi/2)    0
RX(-pi/2)    1
\end{lstlisting}
for a \verb|CZ|-based architecture.

\paragraph{Clifford manipulation} \QUILC{}'s source code includes a comprehensive library for manipulating the Clifford group, including subroutines suitable for randomized benchmarking and stabilizer simulation of Clifford gates of arbitrary dimension. Implementation details of the routines included in \QUILC{} for randomized benchmarking can be found in~\cite{robertels}.

\section{A Chronology of \QUILC{}} \label{sec:history}

\QUILC{} originated as a ``compilation framework'' for \QUIL{} in the summer of 2016. Much of the early work was on the the ``front-end'' scaffolding common to most classical compilers (e.g. control-flow graphs, resource parallelization, and nano-pass concepts \cite{sarkar2005educational}). 

Core work on the modern architecture of this paper began in the summer of 2017 with an implementation of the recursive cosine-sine decomposition of \cite{slepoy2006quantum}. By fall of 2017, the broad structure of the compiler, including the division into separate addressing and compression stages, had been worked out. At this point, duration-based heuristics (with look-ahead) were adopted as the default for \texttt{SWAP} selection, and a number of additional compilation routines (including the optimal 2Q implementation of \cite{shende2004minimal} as well as the recursive quantum Shannon decomposition of \cite{shende2006synthesis}) had been implemented.

In spring of 2018, both parametric compilation and state-aware compilation were introduced. Over the following summer, a number of enhancements to the addressing stage were implemented, including the introduction of additional heuristics (such as A* search for \texttt{SWAP} selection) as well as the adoption of partial logical-to-physical qubit mappings.

On January 30, 2019 \QUILC{} was released as an open source project~\cite{QUILC}. A number of contributions have been made since then. Of relevance to this paper, we mention that work on fidelity-based addressing heuristics continued into fall of 2019, and the compiler front-end was extended to support QASM~\cite{cross2017open} programs in December of that year.
\end{document}